\documentstyle[12pt,titlepage]{article}

\renewcommand{\theequation}{\thesection.\arabic{equation}}

\font\grande=cmr10 scaled \magstep4
\font\medio=cmr10 scaled \magstep2
\outer\def\beginsection#1\par{\medbreak\bigskip
      \message{#1}\leftline{\bf#1}\nobreak\medskip\vskip-\parskip
      \noindent}

\def\laq{\raise 0.4ex\hbox{$<$}\kern -0.8em\lower 0.62 
ex\hbox{$\sim$}}
\def\gaq{\raise 0.4ex\hbox{$>$}\kern -0.7em\lower 0.62 
ex\hbox{$\sim$}}

\begin{document}
\bibliographystyle {unsrt}

\titlepage
\begin{flushright}
CERN-TH/96-87\\
FT/UCM/01/96 \\
gr-qc/9604002
\end{flushright}
\vspace{8mm}
\begin{center}
{\grande Normal Modes for Metric Fluctuations}\\
\vspace{5mm} 
{\grande in a Class of Higher-Dimensional Backgrounds}

\vspace{10mm}

M. Gasperini\footnote{Permanent address: 
{Dipartimento di Fisica Teorica, Un. di Torino, 
Via P. Giuria 1, 10125 Turin,
Italy.}} \\
{\em Theory Division, CERN, CH-1211 Geneva 23, Switzerland} \\
and\\
M. Giovannini\footnote{Permanent address: 
{DAMTP, Silver Street, Cambridge, CB3 9EW, UK.}}
\footnote{E-mail: M.Giovannini@damtp.cam.ac.uk}\\ 
{\em Departamento de Fisica Teorica, Universidad Complutense, 
28040 Madrid, Spain} \\
\end{center}
\vspace{7mm}
\centerline{\medio  Abstract} 

\noindent
\baselineskip=13 pt
We discuss a gauge invariant approach to the theory of cosmological
perturbations in a higher-dimensonal background.  
We find the normal modes which diagonalize the
perturbed action,  for a scalar field minimally coupled 
to gravity, in a higher-dimensional 
manifold ${\cal M}$ of the Bianchi-type I, under the assumption that the 
translations along  an isotropic spatial subsection of
${\cal M}$ are isometries of the full, perturbed background. We show
that, in the absence of scalar field potential, the canonical variables
for scalar and tensor metric perturbations satisfy exactly the same
evolution equation, and we discuss the possible 
dependence of the spectrum on the number of internal dimensions.

\vspace{7mm}
\centerline{{\sl To appear in {\bf Class. Quantum Grav.}}}
\vfill
\begin{flushleft}
CERN-TH/96-87 \\
 March 1996 
\end{flushleft}
\baselineskip=20 pt

\newpage

\renewcommand{\theequation}{1.\arabic{equation}}
\setcounter{equation}{0}
\section{Introduction}

It is well known that the evolution in time of a classical cosmological
background can amplify a given distribution of (initially small)
inhomogeneous fluctuations of the metric and of the matter fields 
\cite{lif}.  If the background is spatially homogeneous and isotropic,   
the fluctuations can be unambiguously classified as scalar, vector and
tensor perturbations, according to their transformation properties
under spatial coordinate transformations on a constant-time
hypersurface (see for instance \cite{br1}). The components of such
perturbations are not invariant under local infinitesimal coordinate
transformations (also called ``gauge" transformations); however, it is
always possible to define variables that are ``gauge-invariant" 
\cite{bardeen}, at least to first order in the perturbation amplitude
(see \cite{4} for a full covariant approach, gauge-invariant to all
orders).

In a cosmological context, the initial conditions for the evolution in
time of the perturbations are naturally provided by the quantum
fluctuations of the metric and of the matter fields in their ground
state \cite{sac}.  The correct normalization of the perturbations to an
initial vacuum fluctuation spectrum, however, requires the
computation of the canonical variables that diagonalize the action
(expanded to second order in the perturbation amplitude), and that
represent the normal modes of oscillation of the matter-gravity
coupled system \cite{br1} (see also \cite{6,7}). Such variables satisfy
canonical commutation relations (or, classically, canonical Poisson
brackets), and determine the absolute magnitude of the two-point
correlation function for the perturbations, in terms of the vacuum
fluctuation amplitude. The definition of these canonical variables is
thus a necessary, preliminary step in order to study the time evolution
of a primordial vacuum perturbation spectrum. 

For a homogeneous and isotropic background, the canonical variable
describing decoupled normal oscillations is known in the case of a
perfect fluid source \cite{8}, and in the case of a scalar field source
\cite{9}. Attempts have been made to extend the canonical treatment
to the case of two coupled scalar fields \cite{10}, but always in the
context of an isotropic metric background.   
The main purpose of this paper is to relax the assumption 
of isotropy of the background manifold
in the computation of the canonical normal modes
of the Einstein-Hilbert action, minimally coupled to a scalar
field source. We are motivated to perform this investigation 
by the development of models of the early Universe based on unified theories 
(such as the superstring theory \cite{11}), where the background manifold is
usually given by the product of a 
$(d+1)$-dimensional ``external" space-time and of an $n$-dimensional 
 ``internal" manifold, whose shape and radial size are possibly
time-dependent. 

In anisotropic backgrounds, a gauge-invariant description of metric
perturbations is complicated by the coupling among modes with
different rotational transformation properties, and relative to
different spatial subsections (for instance,  a coupling between
``external" scalars and ``internal"  tensors becomes 
allowed, in principle).  A possible approach to this problem
\cite{12}, valid when the background $\cal M$ can be factorized as
the product of two (or more) isotropic sub-manifolds ${\cal V}_i$  
(i.e. ${\cal M}= {\cal V}_1 \otimes 
{\cal V}_2 \otimes$....), is to define gauge-invariant
variables in each of the sub-manifolds ${\cal V}_i$. In that context the 
fluctuations may be classified as scalars, vectors and tensors with 
respect to coordinate transformations in the  ``physical" external 
sub-manifold.  A different approach
\cite{13}, which can be applied to any Bianchi-type I 
metric background, is to define gauge-invariant variables with respect
to infinitesimal coordinate transformations defined on the whole
manifold. In that case one obtains a very complicated system of 
perturbed equations which, up to now, has been solved only under the 
simplifying assumptions that the propagation is restricted to an effective 
$(2+1)$-dimensional space-time \cite{14}. 

The approach of the present paper is more similar, in spirit, to that of 
\cite{12}. 
We shall assume that the
translations along the internal dimensions are isometries of the full,
perturbed metric background.  Under this assumption, we shall
compute the canonical variables for the normal oscillations of a
background which is the product of two conformally flat manifolds.
Our result shows that in this case a very simple action, invariant
under global $SU(2) \otimes U(1)$ transformations, can simultaneously account 
for the scalar and tensor linear fluctuations of the metric tensor. 

The assumption that perturbations depend only on the external 
coordinates is justified in the context of a typical Kaluza-Klein 
background describing a phase of dynamical dimensional reduction 
\cite{appe,15a,15b}, in which the accelerated inflationary 
expansion of the external dimensions, with scale factor $a$,  is 
sustained by the accelerated contraction of the internal ones, with 
scale factor $b$. In that case, in fact, the curvature scales of the 
internal and external manifolds are both growing in time: if we accept, 
as it seems natural \cite{15a,15b}, that at the end of the process (at 
a conformal time $\eta_1$) internal and external curvature scale are of 
the same (nearly Planckian) order, i.e. 
$(a_1\eta_1)^{-1}\sim(b_1\eta_1)^{-1}$, then $a\ll b$ for all 
$\eta<\eta_1$. Since $(a/b)^2$ controls the ratio of internal to 
external gradients in the equations determining the time evolution of 
the perturbations, it follows that the dependence on the internal 
coordinates tends to be suppressed for large enough perturbation scales, 
namely for all modes $k$ crossing the horizon well before $\eta_1$, when 
the contribution of internal gradients can be safely neglected. This is 
what happens, in particular, in the scalar field-dominated background 
that we analyze in this paper.

The paper is organized as follows. In Section 2 we present the coupled
evolution equations for the scalar perturbation variables, in a
generalized longitudinal gauge. The equations are directly obtained by
perturbing the equations of motion for the metric and scalar field
background. In Section 3 we expand the perturbed action up to second
order, and we introduce the gauge-invariant variables that reduce the
action to the diagonal, canonical form. As an application of our 
results, we discuss the possible dependence of the tensor perturbation 
spectrum on the number of internal dimensions. 
A brief summary, and our
concluding remarks, are finally presented in Section 4. 

\renewcommand{\theequation}{2.\arabic{equation}}
\setcounter{equation}{0}
\section{Background equations and scalar perturbations}

We start with the $D$-dimensional action for a scalar field, minimally
coupled to gravity:  
\begin{equation}
S= S_{g} + S_{m}= -\frac{1}{6 l_D^2} \int d^D x \sqrt{-g}R +
\int d^D x \sqrt{-g}
\left[\frac{1}{2}g^{\alpha\beta}\partial_{\alpha}\varphi\partial_{\beta}\varphi
- V(\varphi)\right] ,        
\label{action}
\end{equation}
where $l_D= \sqrt{8\pi G_D/3}$ has dimensions of length to the power 
$(D-2)/2$.   
We shall consider a homogeneous, Bianchi-type I metric background,
whose spatial part is the product of two conformally flat manifolds: 
\begin{eqnarray}
g_{\mu\nu}= {\rm diag} \left( a^2(\eta), -a^2(\eta)\delta_{ij},
-b^2(\eta)\delta_{mn}\right) , ~~~~~~~~~~~~ \varphi=\varphi(\eta) , 
\nonumber \\
\mu,\nu = 0,..., D-1=d+n , ~~~~~~ i, j=1,..., d , ~~~~~~ 
m,n = d+1,..., d+n ,
\label{metric}
\end{eqnarray}
and $\eta$ is the
conformal time coordinate (the main results of this paper, however,
can be easily generalized to the case of $d+n$ different scale factors). 
For such background, the equations of motion obtained by varying the
action with respect to $g_{\mu\nu}$ and $\varphi$, 
\begin{equation}
R_\mu^\nu - \frac{1}{2}\delta_{\mu}^\nu R = 3 l_D^2 \left[ 
\partial_\mu\varphi\partial^\nu \varphi -
\frac{1}{2}\delta_\mu^\nu
g^{\alpha\beta}\partial_\alpha
\varphi\partial_\beta\varphi+ \delta_\mu^\nu
V(\varphi)\right] , ~~~~~~
g^{\alpha\beta}\nabla_{\alpha}\nabla_{\beta}\varphi+ \frac{\partial
V}{\partial\varphi}=0 , 
\label{ein}
\end{equation}
reduce simply to 
\begin{eqnarray}
{d(d-1)}{\cal H}^2+{n(n-1)} {\cal F}^2+ 2n d {\cal H}
{\cal F}= 6l_D^2 \left(\frac{\varphi'^2}{2} +  a^2
V\right)~~~\nonumber\\ 
2(d-1) {\cal H}'+{(d-1)(d-2)}{\cal H}^2+2n
{\cal  F}'+{n(n+1)} {\cal F}^2+ 2n (d-2){\cal H}{\cal F}
&=& 6 l_D^2 \left(a^2 V -\frac{\varphi'^2}{2} \right)\nonumber \\
2(n-1){\cal F}'+ 2d {\cal H}'+{d(d-1)} {\cal H}^2+
{n(n-1)}{\cal F}^2+2 (d-1)(n-1) {\cal H}
{\cal F}
&=& 6 l_D^2 \left(a^2 V -\frac{\varphi'^2}{2}\right) 
\nonumber\\
\varphi''+\left[(d-1){\cal H} +n {\cal
F}\right]\varphi'+\frac{\partial V}{\partial\varphi} =0 
,~~~~~~~~~~~~~~ 
\label{background}
\end{eqnarray}
where ${\cal H}= {(\ln{ a})}^{\prime}$, ${\cal F}= {(\ln{b})}^{\prime}$ and 
a prime denotes differentiation with respect to $\eta$. 
These equations are not all independent, and the scalar field 
 equation, for instance,    
 can be obtained from the other Einstein equations. For $\varphi=0$ 
these equations describe a particular Kaluza-Klein vacuum \cite{appe}, 
and are solved by a higher-dimensional generalization of the well-known 
Kasner metric background.

By summing and subtracting the above equations one obtains
\begin{eqnarray}
6 l_D^2 a^2 V &=&  \left[(d-1){\cal H} + n{\cal F}\right]' +
\left[(d-1){\cal H} +n{\cal F}\right]^2 ~ ,  
\nonumber\\
3 l_D^2 \varphi'^2 &=& -\left[(d-1){\cal H} +n{\cal F}\right]' +
 (d-1){\cal H}^2 -n {\cal F}^2 + 2 n {\cal H} {\cal F}~, 
\label{first}
\end{eqnarray}
\begin{equation}
{\cal F}' -{\cal H}' =-( {\cal F} - 
{\cal H}) \left[(d-1){\cal H} +n {\cal F}\right]~~~~.
\label{second}
\end{equation}
We shall discuss in this paper the case in which the 
contribution of the scalar potential  to the background equations is
negligible, which is for instance  a reasonable approximation for the
dilaton-driven phase of string cosmology \cite{15,16}.  
In this case, by setting $V(\varphi)=0$, the combination of eqs. 
(\ref{first}) and (\ref{second}) provides the relation 
\begin{equation}
(\ln {\cal H})'= (\ln{\cal F})' =-
 \left[n{\cal F}+(d-1){\cal H}\right] , 
\label{third}
\end{equation}
which will prove very useful for the computation of the perturbed 
action, in the next section.

We now expand  metric and scalar field perturbations around a solution
of the background equations (\ref{ein}), assuming that all dynamical
variables depend only on the ``external" coordinates $x^i$, $i= 1,...,d$. 
In this case modes with different rotational transformation properties
are decoupled, and the scalar component of the background
perturbations can be written in general as 
\begin{equation}
g_{\mu\nu}\rightarrow g_{\mu\nu}(\eta) + 
\delta g^{(S)}_{\mu\nu}({x}^i, \eta) , ~~~~~~~~~
\varphi \rightarrow \varphi(\eta)+ \chi (x^i, \eta)
\end{equation}
where
\begin{equation}
\delta g_{\mu\nu}^{(S)}= \left(\matrix{2 a^2 \phi&- a^2 B_{i}&0&\cr
- a^2 B_{i}& 2 a^2 \psi\delta_{ij}-2 a^2 E_{ij}&0&\cr
0&0& 2 b^2 \xi\delta_{ab}&\cr}\right)
\label{pert}
\end{equation}
(notations: $B_i=\partial_i B$, $E_{ij}=\partial_i\partial_j E$), and all
variables ($\phi, \psi, \xi, E, B$) depend only on $\eta$ and $x^i$. 
Under an infinitesimal coordinate transformation, preserving the scalar
nature  of the fluctuation \cite{br1}, 
\begin{equation}
x^i\rightarrow \tilde{x}^i = x^i + \partial^i \epsilon (\eta, x^i) , ~~~~~~~
\eta\rightarrow \tilde{\eta} = \eta + \epsilon^0 (\eta, x^i) ,
\end{equation}
the components of scalar perturbations transform as 
\begin{eqnarray}
\phi &\rightarrow & \tilde\phi= \phi - {\cal H} \epsilon^0 - {\epsilon^0}'
\nonumber\\
\psi &\rightarrow & \tilde\psi = \psi  + {\cal H} \epsilon^0
\nonumber\\
\xi & \rightarrow & \tilde\xi =\xi + {\cal F}  \epsilon^0
\nonumber\\
E &\rightarrow &\tilde{E}=E- \epsilon 
\nonumber\\
B &\rightarrow &\tilde{B}=B +\epsilon^0 - \epsilon' 
\nonumber\\
\chi &\rightarrow & \tilde{\chi}= \chi - \varphi' \epsilon^0 ~~~~.
\label{gauge}
\end{eqnarray}
A possible choice of 
``gauge-invariant" (linearly independent) variables is then:
\begin{eqnarray}
\Phi &=& \phi +\frac{1}{a}[(B-E')a]' , 
~~~~~~~~~~
\Psi =\psi -{\cal H}(B-E') ,
\nonumber\\
\Xi &=&\xi - {\cal F}(B-E') , 
~~~~~~~~~~
X = \chi +\varphi' (B-E') .
\label{bardeen}
\end{eqnarray}

We choose in this paper $\tilde{B}=0$ and $\tilde{E}=0$,  which defines
a generalized ``longitudinal"  (or conformally Newtonian) gauge
\cite{br1}, and which leaves 
 the coordinate system completely fixed. By perturbing, in this gauge,
the background equations (\ref{ein}) (with $V=0$) we obtain 
the first-order equations for the classical evolution of the scalar
inhomogeneities. The ($i,j$) component, $i\neq j$, of the perturbed
Einstein equations gives a relation between the three perturbation
variables, 
\begin{equation}
\phi=(d-2)\psi +n\xi. 
\label{condit}
\end{equation}
This allows eliminating $\phi$ everywhere in the perturbation
equations. The ($0,0$) component gives 
\begin{equation}
(d-1) \nabla^2 \psi +n \nabla^2 \xi - \psi' \left[d(d-1){\cal H} + n d{\cal
F}\right]
- n \left[ d{\cal H} + (n-1) {\cal F}\right]\xi'= 3 l_D^2
{\varphi'\chi'} .
\label{I}
\end{equation}
The ($i,i$) components give 
\begin{eqnarray}
(d-1)\psi''+ \psi' \left[ (d-1)(2d-3) {\cal H} + n (2d-3) {\cal F}\right]
+ n \xi'' +
\nonumber\\
+ \xi' \left[ 2 n (d-1){\cal H} + (2n^2+n) {\cal F} -n {\cal H}\right]=
3 l_D^2\varphi'\chi' .
\label{II}
\end{eqnarray}
The ($m, m$) components give
\begin{eqnarray}
d\psi'' +(n-1) \xi'' + \nabla^2 \xi - \nabla^2 \psi +
\psi' \left[2d(d-1) {\cal H} + 2 (d-1)(n-1){\cal F}\right]+
\nonumber\\
+\xi'\left[(d(2n-1) - (n-1)) {\cal H} + 2n(n-1){\cal F}\right]=
3 l_D^2 \varphi'\chi' . 
\label{III}
\end{eqnarray}
The scalar field equation gives
\begin{eqnarray}
\chi''+\left[(d-1){\cal H} +n{\cal F}\right] \chi' - \nabla^2 \chi = 2
\varphi' \left[(d-1) \psi' + n\xi'\right]
\label{dilpert}
\end{eqnarray}
Finally, the ($0, i$) components of the Einstein equations give the
constraint: 
\begin{eqnarray}
(d-1)\psi'+(d-2)\psi \left[(d-1){\cal H} + n{\cal F}\right] + n\xi' 
+n\xi \left[(n+1) {\cal F} + (d-2) {\cal H}\right] =
3l_D^2 \varphi'\chi .
\label{IV}
\end{eqnarray}
All these perturbation equations have been obtained by using the
explicit form (\ref{background}) of the background equations.

It is now convenient to define the variable
\begin{equation}
\lambda = \psi +\frac{n}{d-1}\xi~~ , 
\end{equation}
which satisfies the equation
\begin{equation}
\Box \lambda + 3[(d-1){\cal H} + n{\cal F}]\lambda'=0
\label{lambda}
\end{equation}
($\Box=(\partial/\partial\eta)^2 -\nabla^2$)
obtained by subtracting eq. (\ref{II}) from eq. (\ref{I}). The
combination of eqs.  (\ref{III}) and (\ref{I}) gives
\begin{eqnarray}
 d\left\{\Box\psi+\psi'\left[3(d-1){\cal H}+
\frac{{\cal F}}{d}\left( 2(d-1)(n-1)+nd\right) \right]\right\} = 
\nonumber\\=
-(n-1)\left\{\Box\xi +\xi'\left[ \frac{{\cal H}}{n-1}\left(3dn - d -n
+1\right)+3 n {\cal F}\right]\right\}~~~.
\label{anisotper}
\end{eqnarray}	
By inserting in this equation the expression for 
$\Box\psi$ obtained from (\ref{lambda})  we obtain:
\begin{equation}
\Box\xi +\left[3n{\cal F} + (d-1) {\cal H}\right]\xi'+ 2(d-1){\cal
F}\psi'=0~~~. \label{ultima1}
\end{equation}
Finally, by eliminating $\Box \xi$ by means of the above equation, we
obtain from eq.  (\ref{lambda})
\begin{equation}
\Box\psi+\left[3(d-1){\cal H} +n {\cal F}\right] \psi' +2n {\cal
H}\xi'=0~~~. \label{ultima2}
\end{equation}

The system of equations (\ref{ultima1}) and (\ref{ultima2}) describes
the coupled evolution of the ``external" and ``internal" scalar
perturbation variables, $\psi$ and $\xi$. By contrast, each polarization
mode of tensor perturbations is decoupled from the others. A pure
(transverse, traceless) tensor fluctuation  $h$ of the $d \times d$
external part of the metric background satisfies the free scalar field
equation, minimally coupled to the geometry \cite{17,18}, 
\begin{equation}
{h}''+\left[(d-1){\cal H} + n{\cal F}\right]{h}' -\nabla^2 h=0 , 
\label{tens}
\end{equation}
and is automatically invariant under infinitesimal coordinate
transformations preserving the tensor nature 
of the fluctuations.
The similarities between the time evolution of scalar and
tensor perturbations will become more explicit, however, when 
comparing the canonical variables, which diagonalize the 
action (\ref{action}), expanded up to second order in the amplitude of
the metric fluctuations. This will be done in the next section.

\renewcommand{\theequation}{3.\arabic{equation}}
\setcounter{equation}{0}
\section{Normal modes for canonical oscillations}

In this section we shall expand 
the action (\ref{action}) up to second order in the amplitude of 
scalar fluctuations, and we shall look for the ``normal coordinates",
i.e. for the canonical (gauge-invariant) variables which diagonalize the
perturbed action. To this aim, it is convenient to express the
gravitational part of the action, $S_{gr}$,  in terms of the 
Adler-Deser-Misner (ADM) formalism \cite{ADM}, in such  a way that the
second derivatives of the metric tensor appear only as a total
derivative \cite{fock}. By setting
\begin{equation}
ds^2= (N^2 - N_\alpha N^\alpha)d\eta^2- 2 N_\alpha dx^\alpha d\eta -
\gamma_{\alpha\beta} dx^\alpha dx^\beta~~, 
\label{metricadm}
\end{equation}
where Greek indices (only in this section) run from 1 to $d+n$, we
obtain
\begin{eqnarray}
S_{gr}&=&\frac{1}{6l_D^2} \int d^{D} x
\left[N\sqrt{\gamma}\left(K^{\alpha}_{\beta} K^{\beta}_{\alpha}-
K^2\right)
+\frac{1}{2}\left(\sqrt{\gamma}\gamma^{\alpha\beta}
N\right)_{,\alpha}\left(\ln{\gamma} \right)_{,\beta}
+N_{,\alpha}\left(\sqrt{\gamma}\gamma^{\alpha\beta}
\right)_{,\beta}\right] - \nonumber\\
&-&\frac{1}{6l_D^2} \int d^{D} x \left[
\frac{1}{2}N
{\overline \Gamma}_{\alpha\beta}^{\gamma}\sqrt{\gamma}
\gamma^{\alpha\beta}_
{,\gamma} - {\cal D}_{(1)}\right] , ~~~~~~~~~~~~~~~
\label{admaction}
\end{eqnarray}
where
\begin{eqnarray}
{\cal D}_{(1)} = -2\left(K\sqrt{\gamma}\right)' + 2
\left(\sqrt{\gamma} K N^{\alpha} -
\sqrt{\gamma}\gamma^{\alpha\beta}N_{,\beta}\right)_{,\alpha} -  
\left[N\gamma^{\alpha\beta}
(\sqrt{\gamma})_{,\beta}+(N\sqrt{\gamma}
\gamma^{\alpha\beta})_{,\beta}\right]_{,\alpha} ,\nonumber \\
K_{\alpha\beta}=\frac{1}{2}N^{-1}(\nabla_\alpha N_\beta +
\nabla_\beta N_\alpha- \gamma_{\alpha\beta}') , ~~~~~~~~~~~~~
K=K_{\alpha}^{\alpha}~~~.~~~~~~~~~~~~~~~~  
\label{totale1}
\end{eqnarray}
Here a comma denotes partial differentiation; 
$N_\alpha$ are the shift vectors; $N$ is the lapse function;
 $\gamma^{\alpha\beta}$ is the spatial 
$(d+n)$-dimensional part of the metric tensor; ${\overline \Gamma}$ is
the corresponding Christoffel connection; $\gamma~$=~det
$\gamma_{\alpha \beta}$; $K_\alpha^\beta$ is the
extrinsic curvature of the ($d+n$)-dimensional, $\eta~$=~const
hypersurface. Finally, indices are raised, and covariant derivatives are
computed, with the spatial metric $\gamma_{\alpha \beta}$. 

Comparing eq. (\ref{metricadm}) with the
perturbed form of the metric tensor, eq. (\ref{pert}),  we can easily
express the ADM functions up  to second order in the perturbation
variables as: 
\begin{eqnarray} 
N =a\left(1 +\phi -\frac{1}{2} \phi^2 +\frac{1}{2}B_{i}B_{i}\right) , ~~~~~~
N_{i} =  a^2 B_{i} , ~~~~~~
N_m= 0 ,~~~ \nonumber \\
\gamma^{ij}= a^2(1-2\psi)\delta_{ij}+ 2 a^2 E_{ij} , ~~~~~~~~
\gamma_{mn}= b^2(1-2\xi)\delta_{mn} ,~~~~~~~
\end{eqnarray}
\begin{eqnarray}
\det (\gamma_{\alpha \beta}) =
\det(\gamma_{ij}\otimes\gamma_{mn})= a^{2d} b^{2n} [ 1 &-& 2n \xi - 2d
\psi + 2 E_{ii} + 2n(n-1) \xi^2 +\nonumber\\
+ 4nd\psi\xi - 4nE_{ii} \xi + 2d(d-1) \psi^2 &-& 4(d-1)\psi E_{ii} + 2
E_{ii} E_{jj} - 2 E_{ij}E_{ij}] . 
\label{A}
\end{eqnarray}
By using these definitions in eq. (\ref{admaction}) we obtain the
gravity part of the action, expanded to second order in the
scalar perturbation amplitude. With a similar
technique it is also possible to write the matter part of
the action, up to second order  in the amplitude of the fluctuations of
the scalar field ($\chi$) and of metric.
The result of this long algebraic procedure gives the full-second
order action: 
\begin{eqnarray}
\delta_{(2)} S &=& \delta_{(2)}S_{gr} +
\delta_{(2)}S_{m} =
\frac{1}{6l_D^2} \int d^D x  a^{d-1} b^n \left\{ - d(d-1) {\psi'}^2-
\right.  \nonumber\\
&-& n(n-1)  {\xi'}^2 - 2d(d-1){\cal H} \phi \psi' + (d-1)
\left[(d-2)\psi_{i} - 2\phi_{i}\right]\psi_{i}- 
\nonumber\\
 &-& 2n (n-2) \left[ (n-1){\cal F}
+ d{\cal H} ]\xi\xi' -2n[(n-1){\cal F}+ d {\cal H}\right]\phi\xi' -
2nd\psi'\xi'- 
\nonumber\\
 &-& 2nd\left[ n{\cal F}
+ (d-1){\cal H}\right]\xi\psi' + 2n \xi_{i} \left(2 \psi_{i} + \frac{n-1}{2}
\xi_{i} - \phi_{i}\right)+ 
\nonumber\\
 &+& 6l_D^2\left[(\varphi'(\phi' + d\psi'
+ n\xi')\chi +\frac{1}{2}{\chi'}^2 - \frac{1}{2}\chi_{i}^2\right]
+ 4 (B-E')_{ii} \left[-\frac{n}{2} {\cal F}\xi - \frac{n(d-1)}{2}
{\cal H} \xi -  \right.
\nonumber\\
&-&  \left. \left. 
\frac{ n}{2} \xi'-
 \frac{n}{2} {\cal F}\phi + \frac{nd}{2}{\cal H}\xi - \frac{d-1}{2}
{\cal H} \phi - \frac{d-1}{2} {\cal H} \psi' + \frac{3}{2} l_D^2 \chi
\varphi'\right] \right\}+\frac{1}{6l_D^2} \int d^D x {\cal D}_{(2)} , 
\label{secondorder}
\end{eqnarray}
where the total derivative ${\cal D}_{(2)}$ is given explicitly by
\begin{eqnarray}
{\cal D}_{(2)}&=& \partial_i\left\{a^{d-1} b^n\left[ - 4(d-1){\cal H}
\left(E_{ij}(B-E')_{j} -
E_{jj}(B-E')_{i} + \frac{1}{2}E_{jj}B_{i}\right)+ \right. \right.
\nonumber\\
 &+& (B-E')_{ij} (B-E')_{j} - (B-E')_{jj}(B-E')_{i} + E_{ijl}E_{jl} -
E_{jjl}E_{li} -
\nonumber\\
&-&  2n(n-1) {\cal F} \xi B_{i} - 2n {\cal F} \psi B_{i} - 6 n
{\cal H} \xi B_{i} - n(n-1) {\cal F}^2 (E_{jj} E_{i} -E_{j}E_{ji})-
\nonumber\\
&-&2n\left(d{\cal H} +(n-1){\cal F}\right)\xi B_i + (d-2)\left(n{\cal F} -
2(d-1){\cal H}\right)\psi B_i -6 l_D^2 \varphi'\chi B_i +
\nonumber\\
&+&4n{\cal F} \left((B-E')_{i}  E_{jj} - (B-E')_{j}E_{ji}\right) + 6n{\cal
H}{\cal F}(E_{ji}E_{j} - E_{jj}E_{i}) +
\nonumber\\
&+&  \left. \left. 
d(d-1){\cal H}^2 (E_{ij}E_{j} -E_{jj}E_{i})\right]\right\} +  
\nonumber\\
&+& \partial_\eta\left\{  a^{d-1} b^{n} 
\left[ 2 n (d-2)\left({\cal F} + (d-1){\cal
H}\right) \psi E_{ii} + 2 n d ({\cal H} + {\cal F})\xi E_{ii}
 + 6l^2_D \varphi'\chi E_{ii}  + \right. \right.
\nonumber\\
&+&  2\left((d-1){\cal H}+ n{\cal F}\right) E_{ii} E_{jj} - 
d\left((d-1){\cal H} + n(d-2){\cal F}\right) \psi^2 -
\nonumber\\
&-& \left. \left. 
6 l^2_D \varphi'( \phi + d\psi + n\xi)\chi\right]\right\} .
\label{totale2}
\end{eqnarray}

In order to obtain the previous expressions we have repeatedly used 
the background  equations (\ref{background}) and the relation 
(\ref{third}). The functional 
derivative of the action (\ref{secondorder}) with respect to $(B-E')$ 
provides a constraint \cite{br1} 
which relates the different longitudinal fluctuations: 
\begin{equation}
3l_D^2 \chi\varphi' = \left[(d-1) {\cal H}+ n{\cal F}\right]\phi + (d-1)
\psi' + n\xi'  + n({\cal F}  - {\cal H})\xi~~~. 
\label{constr}
\end{equation}
By using eq. (\ref{condit}) this expression reduces exactly to 
the constraint  (\ref{IV}), which we obtained in the previous section
from the ($0, i$) component of the perturbed 
Einstein equations. This is quite an important consistency check of our 
procedure. 

By inserting eq. (\ref{constr}) in the second-order action
(\ref{secondorder}), we can finally  diagonalize this action by
introducing the following two variables $v$ and $w$:
\begin{equation}
v = a^{\frac{d-1}{2}}b^{\frac{n}{2}}\chi +z\lambda=
a^{\frac{d-1}{2}}b^{\frac{n}{2}} X + z\Lambda ,
\label{v}
\end{equation}
\begin{equation}
w = \frac{z}{l_D}\left[\frac{n(n+d-1)}{6(d-1)}\right]^{1/2} \left(
\frac{{\cal H}}{\varphi'} \xi - \frac{{\cal F}}{\varphi'}\psi\right)=
\frac{z}{l_D}\left[\frac{n(n+d-1)}{6(d-1)}\right]^{1/2}\left( \frac{{\cal
H}}{\varphi'} \Xi - \frac{{\cal F}}{\varphi'}\Psi\right) , 
\label{w} 
\end{equation}
where
\begin{equation}
z=\frac{a^{\frac{d-1}{2}}b^{\frac{n}{2}}\varphi' }{{\cal
H}+\frac{n}{d-1}{\cal F}}  , ~~~~~~~~~~~
\Lambda= \Psi + \frac{n}{d-1} \Xi~~~ .
\label{z}
\end{equation}
The gauge invariance of $w$ and $v$ is simply a consequence of the
gauge invariance of the variables $\Psi$, $X$, $\Xi$, defined in eq.
(\ref{bardeen}).  In terms of $v$ and $w$ the action
(\ref{secondorder}) can be written in the canonical form as the action for
two non-interacting scalar fields, both coupled to the same
time-dependent external potential $z''/z$, namely
\begin{eqnarray}
\delta_{(2)}S= \frac{1}{2}\int d^{D} x\left[ {v'}^2
+\frac{z''}{z}v^2 -v_{i}^2
+ {w'}^2 +\frac{z''}{z}w^2 -w_{i}^2
+2{\cal D}_{(3)} \right]
\label{diagonal}
\end{eqnarray}
where
\begin{eqnarray}
{\cal D}_{(3)}&=& {d\over d \eta}\left\{\frac{d-1}{6l_D^2} \frac{a^{d-1} b^{n}
\lambda_{i}^2 }{{\cal H} +\frac{n}{d-1} {\cal F}}
- \frac{v^2}{2}\left(\frac{d-1}{2}\right)\left({\cal H}+\frac{n}{d-1}{\cal
F}\right) - \frac{3}{d-1}\frac{l_D^2{\varphi'}^2 v^2}{a^{\frac{d-1}{2}}
b^{\frac{n}{2}}}+ \right. \nonumber\\
&+& \frac{3}{2}\frac{l_D^2 z^2
\varphi'}{a^{\frac{d-1}{2}}b^{\frac{n}{2}}}\lambda v-
\frac{n}{d-1}\left[(n-1){\cal F} +d{\cal H}\right] z\xi v +
\frac{n}{d-1}\left[(n-1){\cal F} + d {\cal H}\right] z^2 \lambda \xi
+\nonumber\\ &+& \frac{3}{2(d-1)}
\frac{l_D^2\varphi'}{a^{\frac{d-1}{2}}b^{\frac{n}{2}}}z^3 \lambda^2
+ dzz' (\lambda\psi) - d\psi v z' - \frac{nd}{3}
\frac{a^{d-1}b^n}{l_D^2} \left[(n-1){\cal F} + d{\cal H}\right]
\xi^2+\nonumber\\  &+& \left.
\frac{n^2}{6(d-1)}
\frac{a^{d-1}b^{n}}{l_D^2} \frac{({\cal H}-{\cal F}) [(n-1){\cal F} +
d{\cal H}]}{(d-1){\cal H} +n{\cal F}}\xi^2 \right\}~~~~
\end{eqnarray}
is another total derivative, which does not contribute to the equations
of motion. 

The two variables $v$ and $w$ generalize to the higher-dimensional,
anisotropic case $d>3$, $n\neq 0$, the canonical variable \cite{9} 
representing normal oscillations 
in $d=3$, $n=0$, introduced for a gauge-invariant 
description of scalar perturbations. In the absence of scalar 
field, the equations describing the fluctuations of our 
particular Kaluza-Klein vacuum \cite{appe}
can be directly obtained by setting 
$\varphi = \chi=0$ in the corresponding equations of Section 2.  
In this case there is only one normal mode of oscillation, and
the action reduces to the canonical form (\ref{diagonal}) with $v=0$
and 
\begin{equation}
w=\left[\frac{n(n+d-1)}{6l_D^2(d-1)}\right]^{1/2}\frac{a^{\frac{d-1}{2}}b^
{\frac{n}{2}}}{{\cal
H}+\frac{n}{d-1}{\cal F}} \left( {\cal H} \xi - {\cal F}\psi\right)~~
\label{kaluza}
\end{equation}
(the isotropic, $d$-dimensional case is recovered for $b=$ const, ${\cal
F}=0$). If we have, on the contrary, $N>1$ scalar fields minimally
coupled to gravity,
\begin{equation}
S= -\frac{1}{6 l_D^2} \int d^D x \sqrt{-g}R +
\frac{1}{2}\int d^D x \sqrt{-g}\left[\sum_{k=1}^{N}g^{\alpha\beta}
\partial_{\alpha}\varphi_k\partial_{\beta}\varphi_k \right]        
\end{equation}  
the action (\ref{diagonal}) easily generalizes to the action describing
the oscillations of $N+1$ normal modes, with canonical variables
\begin{equation}
w ,~~~~~~v_k = a^{\frac{d-1}{2}} b^{\frac{n}{2}}\chi_k + z_k \lambda
, ~~~~~~ 
z_k=\frac{a^{\frac{d-1}{2}}b^{\frac{n}{2}}\varphi_k' }{{\cal
H}+\frac{n}{d-1}{\cal F}},~~~~~k =1,..., N
\end{equation}
where $\chi_k=\delta \varphi_k$, and $w$ is the same variable as in 
eq. (\ref{w}). For $n=0$ and $d=3$, this result coincides with the one
recently obtained in \cite{19}. The results of this section cannot 
be directly applied to 
the case in which the 
scalar potential $V(\varphi)$ is non-vanishing, since in 
that case eq. (\ref{third}) is no longer valid and the 
expressions for the total derivative terms become more complicated. 

According to the action (\ref{diagonal}), the Fourier components of the
canonical variables $v$ and $w$ satisfy the evolution equations
\begin{equation}
v_k''+ \left[k^2 - \frac{z''}{z}\right]v_k=0 , ~~~~~~~~~
w_k''+ \left[k^2 - \frac{z''}{z}\right]w_k=0 .
\label{eqw}
\end{equation}
These equations also directly follow from the
definition of $v$ and $w$, and from the evolution
equations of the scalar fluctuations derived in the previous section,
namely eqs. (\ref{dilpert}),  (\ref{ultima1}) and
(\ref{ultima2}). In particular, for a power-like behaviour of the
background, $z(\eta)\sim |\eta|^\alpha$, eqs. (\ref{eqw}) are  
solved exactly by 
\begin{eqnarray}
v_k = \frac{1}{\sqrt{k}}\left[
A_k\sqrt{|k\eta|}H^{(2)}_{\nu}(|k\eta|)
+B_k\sqrt{|k\eta|}H^{(1)}_{\nu}(|k\eta|)\right]~~~~ ,
\nonumber\\
w_k = \frac{1}{\sqrt{k}}\left[
C_k\sqrt{|k\eta|}H^{(2)}_{\nu}(|k\eta|)
+D_k\sqrt{|k\eta|}H^{(1)}_{\nu}(|k\eta|)\right]~~~~,
\label{general}
\end{eqnarray}
where $\nu =|\alpha -1/2|$, and $H_\nu^{(1)}$,  $H_\nu^{(2)}$ are the 
first- 
and second-kind Hankel functions \cite{21}. Once that $v$ and $w$ are
known, the Fourier components of the metric perturbation variables,
$\psi_k$, $\xi_k$, can be expressed in terms of the gauge-invariant
variables as
\begin{eqnarray}
k^2 \psi_{k} &=& \frac{ n(n+d-1) {\cal H}{\cal
F}\varphi'}{[(d-1){\cal H}+ n{\cal
F}]^2}\left[\frac{6l_D^2(d-1)}{n(n+d-1)}\right]^{1/2}
\left(\frac{w_{k}}{z}\right)'-
\nonumber\\
&-& \frac{3l_D^2\varphi' {\cal H}}{[(d-1){\cal H} +n {\cal F}]}
\left(\frac{v_{k}}{a^{\frac{d-1}{2}}b^{\frac{n}{2}}}\right)'
-\frac{n\varphi'}{(d-1){\cal H} + n{\cal F}}
\left[\frac{6l_D^2(d-1)}{n(n+d-1)}\right]^{1/2}
\left(\frac{w_{k}}{z}\right)~ ,
\nonumber\\
\xi_{k} &=& \frac{d-1}{n} (\lambda_{k} - \psi_{k})~~,  \nonumber\\
k^2 \lambda_{k} &=&\frac{n {\cal F}}{d-1} \frac{
[(n+d-1)\varphi']}{[(d-1){\cal H} + n{\cal
F}]}\left[\frac{6 l_D^2(d-1)}{n(n+d-1)}\right]^{1/2}
\left(\frac{w}{z}\right)'
 - \frac{3 l_D^2\varphi'}{(d-1)}\left(\frac{v}{a^{\frac{d-1}{2}}
b^{\frac{n}{2}}}\right)'
\label{lambdav}
\end{eqnarray}
(we have used the whole set of equations (\ref{I})--(\ref{dilpert})). 
For $n=0$, $d=3$, $\xi_k=0$, we recover the standard relation
\cite{br1}
\begin{equation}
\psi_k = - \frac{3l_D^2}{2} \frac{\varphi '}{k^2} 
\left(\frac{v}{a}\right)' . 
 \end{equation}

It is now interesting to compare, at the level of canonical variables,
the behaviour of scalar perturbations with that of tensor
perturbations propagating in a $d=3$ external metric background. In
terms of the two transverse and traceless tensor polarization modes,
$h_\oplus$ and $h_\otimes$, the second-order action for tensor
perturbations, up to a total derivative, can be written 
\begin{equation}
\delta_{(2)} S^{(T)}= {1\over 2} \int d^{4+n}x\left[ u^{'2}_\oplus +
 u^{'2}_\otimes + {y''\over y}(  u^{2}_\oplus + u^{2}_\otimes )
-(\partial_i u_\oplus)^2 -(\partial_i  u_\otimes)^2 \right],
\label{100}
\end{equation}
where the canonical variables $u_\oplus$ and $u_\otimes$ are
defined by \cite{16,17,22}
\begin{equation}
u_{\oplus}=\frac{y}{24l_D}h_{\oplus} ,~~~~~~~
~u_{\otimes}=\frac{y}{24l_D}h_{\otimes} , ~~~~~~~ y = a b^{n/2} . 
\label{322}
\end{equation}
In  the absence of scalar potential we have  
$y''/y= z''/z$. We can thus rewrite
the sum of the two actions (\ref{diagonal}) and (\ref{100}) in 
compact form, invariant under global $SU(2)\otimes U(1)$
transformations: 
\begin{equation}
\delta_{(2)} S^{(S)} + \delta_{(2)} S^{(T)}
=\int d^{4+n}x \left[ \eta^{\mu\nu}(\partial_\mu {\cal
Q})^{\dagger}\partial_\nu {\cal Q} -m^2 {{\cal Q}}^{\dagger} {\cal Q}
\right] , 
\label{finale}
\end{equation}
where 
\begin{equation}
{\cal Q} =  \left(\matrix{q \cr Q \cr}\right), ~~~~
q=\frac{v +iw}{\sqrt{2}} , ~~~~
Q=\frac{u_{\oplus} +iu_{\otimes}}{\sqrt{2}}, ~~~~
m^2= -{z''\over z} =-{y''\over y} ,
\label{bispinore}
\end{equation} 
and $\eta_{\mu\nu}$ is the flat Minkowski metric in $4+n$ dimensions.
This action explicitly displays the similarities in the time evolution 
 of the  gauge-invariant, canonical variables for scalar and
tensor fluctuations,  in spite of the very different equations of motion
for the metric perturbation variables, written in the longitudinal
gauge. This similar behaviour 
was also stressed in \cite{16}, where it was argued that the 
gauge-invariant variables could be more appropriate than the standard
longitudinal ones for a consistent expansion 
of scalar inhomogeneities outside the horizon. 

As a physical application of our results, we shall discuss the possible 
dependence of the tensor perturbation spectrum on the number of internal 
dimensions, by considering a background with $d=3$ and $n\not=0$ (a similar 
discussion can be easily repeated in the case of scalar perturbations). 
The power spectrum $P_h(k)$ is defined as usual \cite{24a} in terms of 
the Fourier transform of the two-point correlation function,
\begin{equation}
\langle h(x_i) h(x_i')\rangle=\int {d^3k\over (2 \pi k)^3} 
e^{ik_i(x_i-x_i')}P_h(k) , 
\label{325}
\end{equation}
where the brackets denote spatial average, or ensamble average over a 
distribution of stochastic variables. For each perturbation mode $h_k$ 
we thus have, from eq. (\ref{322}),
\begin{equation}
P_h(k)\simeq k^3|h_k|^2\simeq {k^3\over M_p^n}{|u_k|^2\over (yM_p)^2} ,
\label{326}
\end{equation}
modulo numerical factors of order unity ($M_p$ is the Planck mass). 
According to the action (\ref{100}), the canonical variable $u_k$ 
satisfies exactly the free evolution equation (\ref{eqw}), with $z$ 
replaced by $y$. In a power-law background with $y\sim |\eta|^\alpha$ 
we thus choose as solution
\begin{equation}
u_k=|\eta|^{1/2}H_\nu^{(2)}(|k\eta|), ~~~~~~~~~~~~~~~
\nu =|\alpha-1/2|,
\label{327}
\end{equation}
which satisfies the correct vacuum normalization \cite{br1} for 
$|k\eta|\gg1$. In the opposite limit $|k\eta|\ll1$ the perturbations 
are amplified by the background evolution, and eq. (\ref{327}) gives 
$|u_k|\simeq |k\eta|^\alpha/\sqrt k=y/(\sqrt k y)_{hc}$, where ``$hc$" 
denotes the time of horizon crossing, $\eta=k^{-1}$ (we have assumed 
$\alpha <1/2$, corresponding to the case in which the comoving amplitude 
of perturbations stay frozen outside the horizon; for 
a discussion of the opposite case 
see for instance \cite{24b}). Since $y^2=a^2b^n$ in $d=3+n$, this gives 
the normalized spectrum as
\begin{equation}
P_h(k)\simeq M_p^{-n}\left (y^{-2}\right)_{hc}\left(k\over M_p\right)^2 
\simeq \left(k\over a M_p\right)^2_{hc} \left(bM_p\right)^{-n}_{hc}
\simeq \left(H\over  M_p\right)^2_{hc} \left(bM_p\right)^{-n}_{hc}
\label{328}
\end{equation}
($H=d(\ln a)/dt$ is the usual Hubble parameter defined with respect to 
cosmic time $t$).

When $n=0$ we thus recover the standard tensor perturbation spectrum, 
determined by the Hubble factor at horizon crossing; when $n\not=0$ the 
spectrum 
seems instead to be affected by the dynamics of the internal dimensions. 
For the scalar field-dominated background of this paper, however, the 
modification is only apparent, because from the general solution of the 
background equations \cite{15} we have $y^2=a^2b^n\sim |\eta|$, i.e. 
$\alpha =1/2$. The $n$-dependence disappears from $y$ so that, according 
to eq. (\ref{328}), $P_h(k)\sim k^{2+2\alpha}\sim k^3$ (modulo 
logarithmic corrections) like in a 
four-dimensional background \cite{15,16}, quite independently of $n$. 

The situation is obviously different when other sources are present, 
beside the scalar field. Consider, for instance, the addition of perfect 
fluid matter, with energy density $\rho$ and pressure $p_1=\gamma_1 
\rho$, $p_2=\gamma_2 \rho$, in the external and internal submanifolds, 
respectively. Tensor perturbations are decoupled from the fluid 
sources, so that  
the canonical variables are the same, and the normalized spectrum is 
still given by eq. (\ref{328}). From the general solution of the 
background equations with perfect fluids, given in \cite{15}, we get 
(in the Einstein frame):
\begin{equation}
y^2=a^2b^n\sim |\eta|^{2\alpha} , ~~~~~~~~~~~~~~
2\alpha={2(1-\gamma_1)\over 1-2\gamma_1+3\gamma_1^2+n\gamma_2^2} .
\label{329}
\end{equation}
The spectrum (\ref{328}), $P_h(k)\sim k^{2+2\alpha}$, is now 
$n$-dependent (unless we put $\gamma_2=0$), and the contribution of the 
internal dimensions goes in the direction of flattening the spectrum 
with respect to the pure $d=3$ case. 

\renewcommand{\theequation}{4.\arabic{equation}}
\setcounter{equation}{0}
\section{Discussion and conclusion}

In this paper we 
discuss a Lagrangian approach to the perturbations 
of scalar  field matter, minimally coupled to gravity in a homogeneous 
cosmological background.  
We explicitly consider a metric of the Kaluza-Klein
type, corresponding to the direct product of two conformally flat
(internal and external) manifolds. Under the assumption of frozen
dependence on the internal coordinates (namely for perturbations only
depending on the variables of one of the two spatial sub-manifolds), 
we find the canonical variables that diagonalize the
original action, up to second order in the amplitude of the 
metric and scalar field fluctuations.  The total action can then be
written in a compact form in terms of a complex bivector, which
provides a unified description of scalar and tensor normal 
 excitations of the background manifold. 

This suggests that, 
in a string cosmology context, the
$O(d+n,d+n)$ covariance of the background equations \cite{23} should
characterize not only the evolution of tensor perturbations \cite{24},
but that of scalar perturbations (in the String frame) as well. In the 
Einstein frame, used in this paper, it is interesting to observe that 
the time evolution of the scalar and tensor canonical variables is 
determined by the background function
\begin{equation}
{z''\over z}={y''\over y}=-\left(d-1\over 2\right)^2\left( {\cal H}
+{n {\cal F}\over d-1}\right)^2= -\left(z'\over z\right)^2 ~~,
\end{equation}
which is invariant under the transformation $a \rightarrow a^{-1}$, 
$b \rightarrow b^{-1}$, which implies $z \rightarrow -z^{-1}$. It is 
also invariant, separately, under the transformation $\varphi 
\rightarrow -\varphi$, which implies $z \rightarrow -z$. In the Einstein 
frame this second transformation corresponds indeed, in the absence of 
potential for the scalar field, to a scale factor duality transformation 
\cite{27} for an isotropic, dilaton-driven background, evolving in time 
with a power-like behaviour \cite{15}.  If the background admits, in 
particular, an asymptotic regime in which the Laplacian terms become 
negligible in the evolution equations, then the invariance of $z''/z$ implies 
the invariance, in that regime,  of $q''/q$ and ${ Q}''/{ Q}$ as a functions 
of $z$.  

As already mentioned in the Introduction, and stressed in \cite{br1,6,7},
the correct normalization of the metric perturbations to an initial
vacuum fluctuation spectrum is only possible after introducing 
the gauge-invariant variables representing the 
normal modes, which
diagonalize the action and satisfy canonical commutation relations.
This paper should be regarded as a first step towards the definition 
of such variables in the case of higher-dimensional backgrounds. It
would be important, however, to drop the assumption that
fluctuations depend only on the external coordinates. The dependence
on internal coordinates modifies in fact the perturbation equations: for
the tensor case, by adopting for instance the approach of \cite{12},
one obtains 
\begin{eqnarray}
{h_i^j}'' + [(d-1){\cal H} + n{\cal F}] {h_i^j}' -\nabla^2 _{{x}} {h_i^j} -
\frac{a^2}{b^2} \nabla^2_{{y}} {h_i^j} =0 ,
\nonumber\\
{h_a^b}'' + [(d-1){\cal H} + n{\cal F}] {h_a^b}' -\nabla^2 _{{x}} h_a^b -
\frac{a^2}{b^2} \nabla^2_{{y}} h_a^b =0 , 
\label{pol}
\end{eqnarray}
where $\nabla^2_{{x}}$ and $\nabla^2_{{y}}$ are, respectively,
the external and internal Laplacian operator; $h_i^j$ and  $h_a^b$ are
the external and internal polarization modes. Unless the terms with
the internal Laplacian, asymptotically, become sub-leading with respect
to the others, the 
dependence on the internal coordinates can modify in a significant
way the power spectrum computed under the assumption that
such coordinates are frozen.

Finally, it would be interesting to repeat the analysis of this paper in
the context of the Hamiltonian formalism, more  
appropriate for the discussion of the constraints that characterize
scalar perturbations, and for the standard approach to the canonical
quantization procedure of the fluctuations.

\section*{Acknowledgements}

We are grateful to  Gabriele Veneziano for many useful
discussions. M. Giovannini wishes to thank the CERN Theory
Division for hospitality  during the completion of this research. This 
work is supported in part by the ``Human Capital and Mobility
Program" of the European Commission, under the contracts No.
CHRX-CT94-0423 (M. Giovannini) and No. ERBCHRX-CT94-0488 
(M. Gasperini).

\end{document}